\documentclass[aps,pra,reprint,superscriptaddress,nofootinbib]{revtex4-1}
\usepackage{amsmath}
\usepackage{amssymb}
\usepackage{amsfonts}
\usepackage{mathrsfs}
\usepackage{graphicx}
\usepackage[caption=false]{subfig}
\usepackage{enumitem}
\usepackage[normalem]{ulem}
\usepackage[percent]{overpic}
\usepackage{bm}
\usepackage{rotating}
\usepackage[usenames, dvipsnames]{color}
\usepackage{hyperref}
\usepackage{cancel}
\usepackage{verbatim}
\usepackage{mathtools}
\usepackage{graphicx}
\usepackage{tensor}
\usepackage{verbatim}
\usepackage{MnSymbol}
\usepackage{booktabs}
\graphicspath{ {images/} }

\usepackage{braket}

\newcommand{\ii}{\mathrm{i}}

\newcommand{\be}{\begin{equation}}
\newcommand{\bel}[1]{\begin{equation}\label{#1}}
\newcommand{\ee}{\end{equation}}

\begin{document}
\title{Zeno Friction and Anti-Friction from Quantum Collision Models}
\author{Daniel Grimmer}
\email{dgrimmer@uwaterloo.ca}
\affiliation{Institute for Quantum Computing, University of Waterloo, Waterloo, ON, N2L 3G1, Canada}
\affiliation{Dept. Physics and Astronomy, University of Waterloo, Waterloo, ON, N2L 3G1, Canada}

\author{Achim Kempf}
\affiliation{Dept. Applied Math., University of Waterloo, Waterloo, ON, N2L 3G1, Canada}
\affiliation{Dept. Physics and Astronomy, University of Waterloo, Waterloo, ON, N2L 3G1, Canada}
\affiliation{Institute for Quantum Computing, University of Waterloo, Waterloo, ON, N2L 3G1, Canada}
\affiliation{Perimeter Institute for Theoretical Physics, Waterloo, ON, N2L 2Y5, Canada}

\author{Robert B. Mann}
\email{rbmann@uwaterloo.ca}
\affiliation{Dept. Physics and Astronomy, University of Waterloo, Waterloo, ON, N2L 3G1, Canada}
\affiliation{Institute for Quantum Computing, University of Waterloo, Waterloo, ON, N2L 3G1, Canada}
\affiliation{Perimeter Institute for Theoretical Physics, Waterloo, ON, N2L 2Y5, Canada}

\author{Eduardo Mart\'{i}n-Mart\'{i}nez}
\email{emartinmartinez@uwaterloo.ca}
\affiliation{Institute for Quantum Computing, University of Waterloo, Waterloo, ON, N2L 3G1, Canada}
\affiliation{Dept. Applied Math., University of Waterloo, Waterloo, ON, N2L 3G1, Canada}
\affiliation{Perimeter Institute for Theoretical Physics, Waterloo, ON, N2L 2Y5, Canada}

\begin{abstract}
We  analyze the quantum mechanics of the  friction experienced by a small system as it moves non-destructively with velocity $v$ over a surface. Specifically, we model the interactions between the system and the surface with a \textit{collision model}. We show that, under weak assumptions, the magnitude of the friction induced by this interaction decreases as $1/v$ for large  velocities. Specifically, we predict that this phenomenon occurs in the Zeno regime, where each of the system's successive couplings to subsystems of the surface is very brief. In order to investigate the friction at low velocities and with velocity-dependent coupling strengths, we motivate and develop \textit{one-dimensional convex collision models}. Within these models, we  obtain an analytic expression for the general friction-velocity dependence.  We are thus able to determine exactly the conditions under which the usual friction-velocity dependency arises. Finally, we give examples that demonstrate the possibility, in principle, of anti-friction, in which case the system is accelerated by its interaction with the surface, a phenomenon associated with active materials and inverted level populations.
\end{abstract}

\maketitle

\section{Introduction}

The study of friction has a very long history, dating back to Aristotle, Vitruvius, and Pliny the Elder \cite{chatterjee2011tribological}, with the first systematic investigations dating back to da Vinci and   Amontons \cite{Popova2015}. 

Today, much is known about both classical and quantum aspects of friction. In particular, the term ``quantum friction'' has been used for a phenomenon that has its origin in the fact that even neutral objects interact through the vacuum of the quantized electromagnetic field, namely through the Casimir and Van der Waals forces.
``Quantum friction'' then is the phenomenon that a neutral quantum system that moves near a stationary object, for instance an atom moving over a plate, may experience van der Waals or Casimir-type forces that also possess a component directed against the system's motion, see, e.g., \cite{Milton_2016,Intravaia_2015,PhysRevA.94.042114,PhysRevA.95.052510,Barton_2010,Høye2014,PhysRevA.98.032507}.

In contrast, in the present paper, we 
investigate the emergence of friction when a generic quantum system moves non-destructively over a surface while interacting directly (i.e., not mediated by a quantum field) with the individual subsystems of the surface that it encounters on its way. We model the effective interaction of the moving particle with every constituent of the surface through a \textit{Collision Model}.

Within this model, we recover the expected velocity dependence of the friction force for small velocities. Quite counter-intuitively, under mild assumptions, we find that for large  velocities the friction decays as $1/v$, a phenomenon we refer to as Zeno friction. We also find that ``anti-friction" is possible under certain circumstances.

\section{Friction in Collision Models}

\begin{figure}
\includegraphics[width=0.35\textwidth]{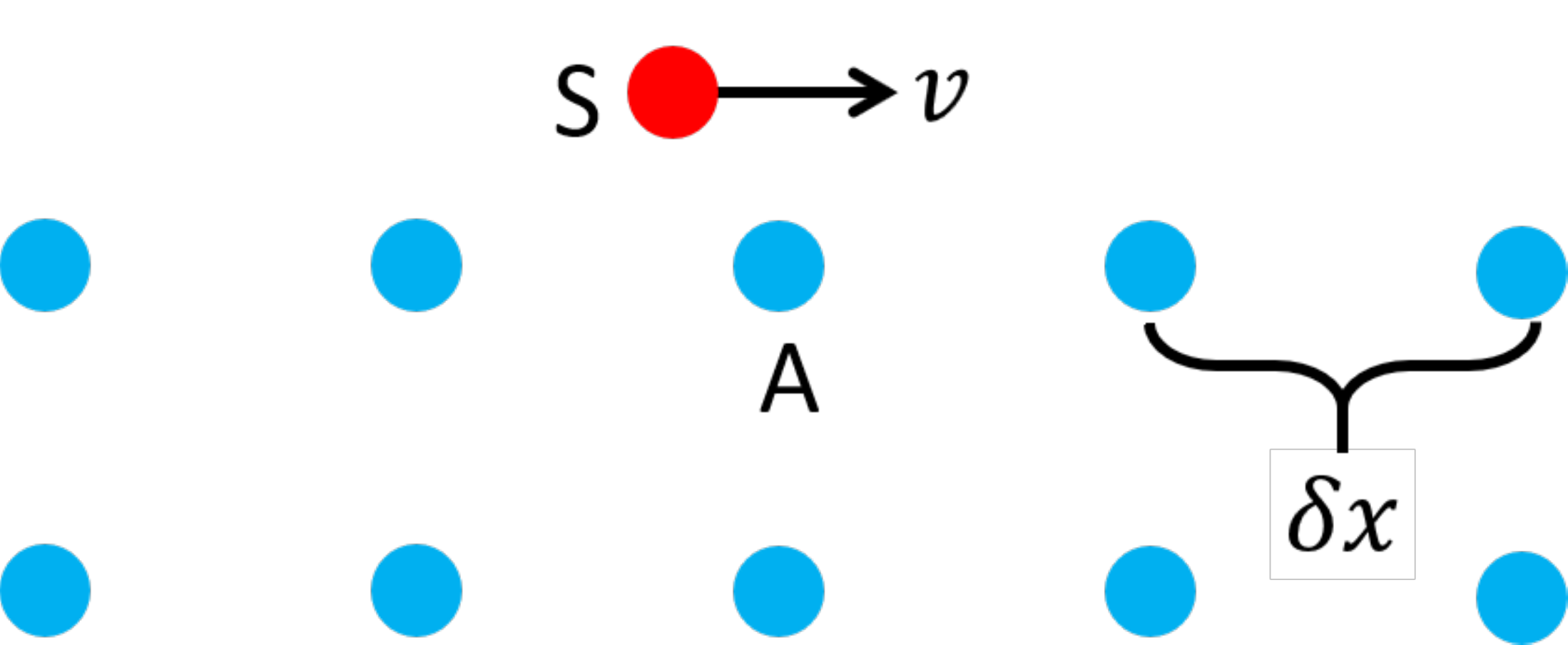}
\caption{(Color online.) A quantum system, $S$, being pulled at a fixed speed, $v$, by an external agent over a line of ancillary systems. The ancillas are separated by a distance $\delta x$. We assume that the system only interacts with the nearest ancilla, $A$, at any given moment.}
\label{CollisionFigure}
\end{figure}

Consider a quantum system, $S$, pulled by an external agent at a fixed speed $v$ over a line of ancillary systems, as in Fig. \ref{CollisionFigure}. Suppose that these ancillas are separated by a distance $\delta x$ and that the system only interacts with the nearest ancilla at any given moment such that the system interacts with a new ancilla, $A$, every $\delta t=\delta x/v$. 

During the moving system's interaction with an ancilla, the two systems will exchange energy between their internal degrees of freedom.  However, the total internal energy of the moving system and the ancilla system is generally not conserved. This is because an external agent generally has to apply a force to maintain the chosen velocity of the moving system. The work $\delta W$ done\footnote{Note that the work associated with each interaction is defined in terms of the expected energy changes of the system and ancilla. Generally, the work required to perform a quantum process is associated with a distribution of work costs \cite{RevModPhys.83.771}. In general, these distributions can have variances comparable to their averages. An analysis of the quantum fluctuations of this work cost (and of the friction we define from it) would be interesting but is beyond the scope of this paper. We do note however, that the interactions we consider are applied repeatedly such that fluctuations of the individual interactions will tend to cancel out and make the total work cost relatively more certain.} by the agent per object-ancilla interaction satisfies the relation
\be
\delta E_\text{S}
+\delta E_\text{A}
+\delta W=0,
\ee
where $\delta E_\text{S}$ and $\delta E_\text{A}$ are the changes in the system and ancilla's internal energies due to the interaction.

This energy cost of maintaining a fixed speed per distance travelled is due to friction which,  time-averaged over the object-ancilla interaction, reads: 
\begin{align}\label{FrictionDef}
f&\coloneqq -\frac{\delta W}{\delta x}
=\frac{\delta E_\text{S}}{\delta x}
+\frac{\delta E_\text{A}}{\delta x}
=\frac{1}{v}\Big(\frac{\delta E_\text{S}}{\delta t}
+\frac{\delta E_\text{A}}{\delta t}\Big).
\end{align}
Note that in our sign convention the friction is positive if the agent that keeps the object at a constant speed continually loses energy to the system and ancilla.  


If there is no external agent, i.e., if the system moves over the ancillas freely, then the energetic cost of the friction is paid by the moving system's kinetic energy. Except, as we will show below, there can occur scenarios of anti-friction, in which case the moving object would speed up, or push against whatever might hold it back. 

Concretely, we will model and analyze the friction and antifriction discussed above using the framework of {Collision Models}. Counting from $n=0$, in the $n^{th}$ interaction the states of the  system and ancilla update as
\begin{align}\label{GeneralUpdateS}
\rho_\text{S}(n \, \delta t)&
\to\text{Tr}_\text{A}(U(\delta t)(\rho_\text{S}(n\,\delta t)\otimes\rho_\text{A}(0))U(\delta t)^\dagger)\\
\nonumber
&=\Phi_\text{S}(\delta t)[\rho_\text{S}(n \, \delta t)],\\
\label{GeneralUpdateA}
\rho_\text{A}(0)
&\to\text{Tr}_\text{S}(U(\delta t)(\rho_\text{S}(n\,\delta t)\otimes\rho_\text{A}(0))U(\delta t)^\dagger)\\
\nonumber
&=\Phi_\text{A,n}(\delta t)[\rho_\text{A}(0)],
\end{align}
where $U(\delta t)$ is some unitary operator on the joint system, $SA$, describing their interaction. Note that the ancilla always starts in the state $\rho_\text{A}(0)$. Also note while the system's update map, $\Phi_\text{S}(\delta t)$, is independent of $n$, the ancilla's update map, $\Phi_\text{A,n}(\delta t)$, can depend on the interaction number, $n$, through the system's current state, $\rho_\text{S}(n \, \delta t)$. Further note that $\Phi_\text{A,n}(\delta t)$ always depends on $\rho_\text{S}(n\,\delta t)$ linearly.

From these update formulae  we can compute the expected change in the system's internal energy  
\be
\delta E_\text{S,n}
=\text{Tr}_\text{S}(\hat{H}_\text{S} \, (\Phi_\text{S}(\delta t)-\openone_\text{S})[\rho_\text{S}(n\,\delta t)]),
\ee
where $\hat{H}_\text{S}$ is the local Hamiltonian of $S$ and $\openone_\text{S}$ is the identity channel on $S$. Likewise we can compute the expected change in the ancilla's internal energy  
\be
\delta E_\text{A,n}
=\text{Tr}_\text{A}(\hat{H}_\text{A} \, (\Phi_\text{A,n}(\delta t)-\openone_\text{A})[\rho_\text{A}(0)]),
\ee
where $\hat{H}_\text{A}$ is the local Hamiltonian of $A$ and $\openone_\text{A}$ is the identity channel on $A$. From these we can identify the average friction during the $n^{th}$ interaction,
\begin{align}\label{FrictionDefN}
f_n
=\frac{\delta E_\text{S,n}}{\delta x}
+\frac{\delta E_\text{A,n}}{\delta x}
=\frac{1}{v}\Big(\frac{\delta E_\text{S,n}}{\delta t}
+\frac{\delta E_\text{A,n}}{\delta t}\Big).
\end{align}
As we will now see, under some natural assumptions, this collisional model of friction yields unexpected--and even  bizarre--phenomenology at high speeds.

\subsection{Collisional Friction in the Zeno Regime}\label{SecZenoFriction}

It is often natural to expect that nothing can happen in no time and that when things do happen they happen at a finite rate. We can capture these intuitions by making some regularity assumptions about the update maps' behaviors around $\delta t=0$. Specifically, we could assume that
\bel{Regularity1}
\Phi_\text{S}(\delta t)\to\openone_\text{S}
\quad\text{and}\quad
\Phi_\text{A,n}(\delta t)\to\openone_\text{A}
\quad\text{as}\quad
\delta t\to 0,
\ee
and that,
\bel{Regularity2}
\Phi_\text{S}'(0)
\quad\text{and}\quad
\Phi_\text{A,n}'(0)
\quad\text{exist}
\ee
where the primes indicate a derivative with respects to $\delta t$. For instance, these assumptions hold if the unitary matrix, $U(\delta t)$, in \eqref{GeneralUpdateS} and \eqref{GeneralUpdateA} describing the interaction between $S$ and $A$ are generated by a Hamiltonian, $\hat{H}$, which is independent of $v$ (and therefore of $\delta t$). That is, \mbox{$U(\delta t)=\exp(-\ii \, \hat{H} \delta t/\hbar)$}.

Given these regularity assumptions, it follows that the friction decays as $1/v$ as $v\to\infty$. Specifically taking the limit $v\to\infty$ (or equivalently $\delta t\to0$) in \eqref{FrictionDefN} we find,
\begin{align}\label{ZenoFriction}
f_n&=\frac{1}{v}\text{Tr}_\text{S}(\hat{H}_\text{S} \, \Phi_\text{S}'(0)[\rho_\text{S}(n\,\delta t)])\\
\nonumber
&+\frac{1}{v}\text{Tr}_\text{A}(\hat{H}_\text{A} \, \Phi_\text{A,n}'(0)[\rho_\text{A}(0)])
+\text{o}(v^{-1}),
\end{align}
for large $v$. Note that we are using small-o notation here since we have not assumed $\Phi_\text{S}(\delta t)$ and $\Phi_\text{A,n}(\delta t)$ are second differentiable at $\delta t=0$. This means we see less friction as we go faster. This goes against a common intuition that friction is a penalty for going fast -- in Zeno friction, we see no friction.

As a concrete example, suppose that the unitary matrix governing the interaction is given by
\be
U(\delta t)=\exp(-\ii \,  \hat{H} \, \delta t/\hbar)
\ \ \text{where}\ \   \hat{H}=\hat{H}_\text{S}+\hat{H}_\text{A}+\hat{H}_\text{SA}
\ee
with $\hat{H}$ independent of the systems' relative velocity, $v$. In this case we can easily compute $\Phi_\text{S}'(0)$ and $\Phi_\text{A,n}'(0)$ (as in \cite{Layden:2015b}). From \cite{Layden:2015b} we have
\begin{align}
\Phi_\text{S}'(0)[\rho_\text{S}]
&=\frac{-\ii}{\hbar}[\hat{H}_\text{S}+\text{Tr}_\text{A}(\hat{H}_\text{SA}\,\rho_\text{A}(0)),\rho_\text{S}]\\
\nonumber
&=\frac{-\ii}{\hbar}\text{Tr}_\text{A}\big([\hat{H}_\text{S}+\hat{H}_\text{SA},\rho_\text{S}\otimes\rho_\text{A}(0)]\big).
\end{align}
From this we can compute the first term in \eqref{ZenoFriction} to be
\begin{align}
&\frac{1}{v}\text{Tr}_\text{S}(\hat{H}_\text{S} \, \Phi_\text{S}'(0)[\rho_\text{S}(n\,\delta t)])\\
\nonumber
&= -\frac{1}{v}\frac{\ii}{\hbar}\text{Tr}_\text{SA}(\hat{H}_\text{S} \, [\hat{H}_\text{S}+\hat{H}_\text{SA},\rho_\text{S}\otimes\rho_\text{A}(0)])\\
\nonumber
&=-\frac{1}{v}\frac{\ii}{\hbar}\text{Tr}_\text{SA}([\hat{H}_\text{S},\hat{H}_\text{S}+\hat{H}_\text{SA}]\,\rho_\text{S}\otimes\rho_\text{A}(0))\\
\nonumber
&=-\frac{1}{v}\frac{\ii}{\hbar}\langle[\hat{H}_\text{S},\hat{H}_\text{SA}]\rangle_n
\end{align}
where we have used the identity $\text{Tr}(A[B,C])=\text{Tr}([A,B]C)$ and defined $\langle\,\cdot \,\rangle_n$ as the expectation value taken with respects to the joint state at $t=n\,\delta t$, that is $\rho_\text{S}(n\,\delta t)\otimes\rho_\text{A}(0)$. The second term in \eqref{ZenoFriction} can be computed  by the same method to be \mbox{ $-\frac{1}{v}\frac{\ii}{\hbar}\langle[\hat{H}_\text{A},\hat{H}_\text{SA}]\rangle_n$}. Thus in total the friction is 
\begin{align}\label{ZenoFrictionLayden}
f_n&=\frac{1}{v}\Big\langle\frac{\ii}{\hbar}[\hat{H}_\text{SA},\hat{H}_\text{S}+\hat{H}_\text{A}]\Big\rangle_n+\mathcal{O}(v^{-2}).
\end{align}
Note that as expected the presence of friction is directly related to the non-conservation of the systems' local energies under the interaction Hamiltonian.

This phenomena of decreasing friction at higher velocities is not the sort of velocity dependence that we are used to seeing in our everyday encounters with friction; typically the amount of friction either increases or stays constant at increasing speeds.  One is led to wonder: at what speeds do we expect to start seeing Zeno friction?

To estimate the speeds associates with   Zeno friction, let us consider a particle travelling through the air at a speed $v$ interacting with nitrogen molecules via a Van der Waals interaction (with energy scale $E=10^{-20}\text{ J }=62\text{ meV }=95\,\hbar\text{ THz}$) as it crosses their Van der Waals radius ($r=0.23\text{ nm}$). The perturbative expansion underlying \eqref{ZenoFriction} and \eqref{ZenoFrictionLayden} requires that the amount of evolution happening in each interaction is small, $\delta t \, E/\hbar\ll1$. Taking the duration of the interaction to be the crossing time, \mbox{$\delta t=2 \, r/v$}, we find this requires,
\bel{ZenoCriticalSpeed}
v\gg\frac{2 \, r \, E}{\hbar}
=43 \, \text{km/s}
=1.5\times10^{-4} \, c.
\ee

An important caveat to our prediction of Zeno Friction at high velocities is that the interaction must obey the regularity assumptions, \eqref{Regularity1} and \eqref{Regularity2}. These can be naturally negated by taking the coupling strength between $S$ and $A$ to increase with their relative velocity. For example, if $U(\delta t)$ is generated by $\hat{H}=v \, \hat{H}_0$ then $U(\delta t)\to\exp(-\ii \,  \hat{H}_0 \, \delta x/\hbar)$ as $\delta t\to0$; That is, something happens in no time. Such velocity dependent couplings could arise naturally if the systems couple to each others external/kinetic degrees of freedom. 



Barring this possibility, we expect to see Zeno friction at high velocities. That is, we predict that for velocity independent couplings the amount of friction will begin decreasing at high enough speeds. 

In order to explore friction at low velocities (outside of the Zeno regime) and the possibility of velocity dependent couplings we will now particularize to a simplified class of collision models.  Using these models we be able to reproduce the common friction-velocity profiles we experience everyday. We will also explore scenarios within this model exhibiting anti-friction.

\section{One-dimensional convex collision models}
\subsection{Motivation and Definition}\label{MotivatingExamples}
One of the most widely used collision models \cite{PhysRevLett.113.100603,PhysRevA.75.052110,PhysRevE.97.022111,1367-2630-16-9-095003,PhysRevA.76.062307} is the partial swap interaction first described in \cite{Scarani2002}. It consists of a system, S, interacting with an ancilla, $A$, via the partial swap Hamiltonian, $\hat{H}_\text{sw}=\hbar \, J \, U_\text{sw}$, where $U_\text{sw}$ is the unitary matrix which swaps the states of $S$ and $A$ as \mbox{$U_\text{sw}(\ket{S}\otimes\ket{A})=\ket{A}\otimes\ket{S}$}. Note that $U_\text{sw}$ is self-adjoint, $U_\text{sw}^\dagger=U_\text{sw}$, as well as unitary such that $U_\text{sw}^2=\hat{\openone}_\text{SA}$.  For example, if $S$ and $A$ are qubits then \mbox{$\hat{H}_\text{sw}=\hbar \, J(\hat{\openone}_\text{SA}+ \hat{\bm{\sigma}}_\text{S}\cdot\hat{\bm{\sigma}}_\text{A})/2$} is the isotropic spin coupling. 

Evolution under the partial swap Hamiltonian for a time $t$ is described by the partial swap unitary,
\begin{align}
U(t) 
&=\exp(-\ii \, \hat{H}_\text{sw} \, t/\hbar)\\
&=\cos(J\,t) \, \hat{\openone}_\text{SA}
-\ii \, \sin(J\,t) \, U_\text{sw},
\end{align}
where $\hat{\openone}_\text{SA}$ is the identity operator on the joint system $SA$. Evolving by this unitary from an initially uncorrelated state, the reduced state of the system is,
\begin{align}
\rho_\text{S}(t)
&=\text{Tr}_\text{A}(U(t)(\rho_\text{S}(0)\otimes\rho_\text{A}(0))U(t)^\dagger)\\
\nonumber
&= \cos(Jt)^2 \ \rho_\text{S}(0)
+ \sin(Jt)^2 \ \rho_\text{A}(0)\\
\nonumber
&-\ii \, \cos(Jt) \sin(Jt) \ \text{Tr}_\text{A}([U_\text{sw},\rho_\text{S}(0)\otimes\rho_\text{A}(0)]).
\end{align}
A similar expression holds for the reduced state of the ancilla. The cross terms in these expressions vanish if $\rho_\text{S}(0)$ and $\rho_\text{A}(0)$ are diagonal in the same\footnote{``Same'' here meaning that \mbox{$U_\text{sw}(\rho_\text{S}(0)\otimes\hat{\openone}_\text{A})U_\text{sw}^\dagger=\hat{\openone}_\text{S}\otimes\rho_\text{S}(0)$} and 
$\hat{\openone}_\text{S}\otimes\rho_\text{A}(0)$ are diagonal in the same basis.} basis yielding,
\begin{align}
\rho_\text{S}(t)
&= \cos(Jt)^2 \ \rho_\text{S}(0)
+ \sin(Jt)^2 \ \rho_\text{A}(0),\\
\rho_\text{A}(t)
&= \cos(Jt)^2 \ \rho_\text{A}(0)
+ \sin(Jt)^2 \ \rho_\text{S}(0).
\end{align}
That is, the system and ancilla oscillate between their own initial states and the other's initial state at a rate $J$. Note that each system evolves within a one-dimensional space as a convex combination of two fixed endpoints. We can regard this evolution as the information about the system's initial condition is being passed from S to A and back in the same way that a harmonic oscillator passes its energy between its position (potential energy) and momentum (kinetic energy).

\begin{figure*}
\includegraphics[width=0.32\textwidth]{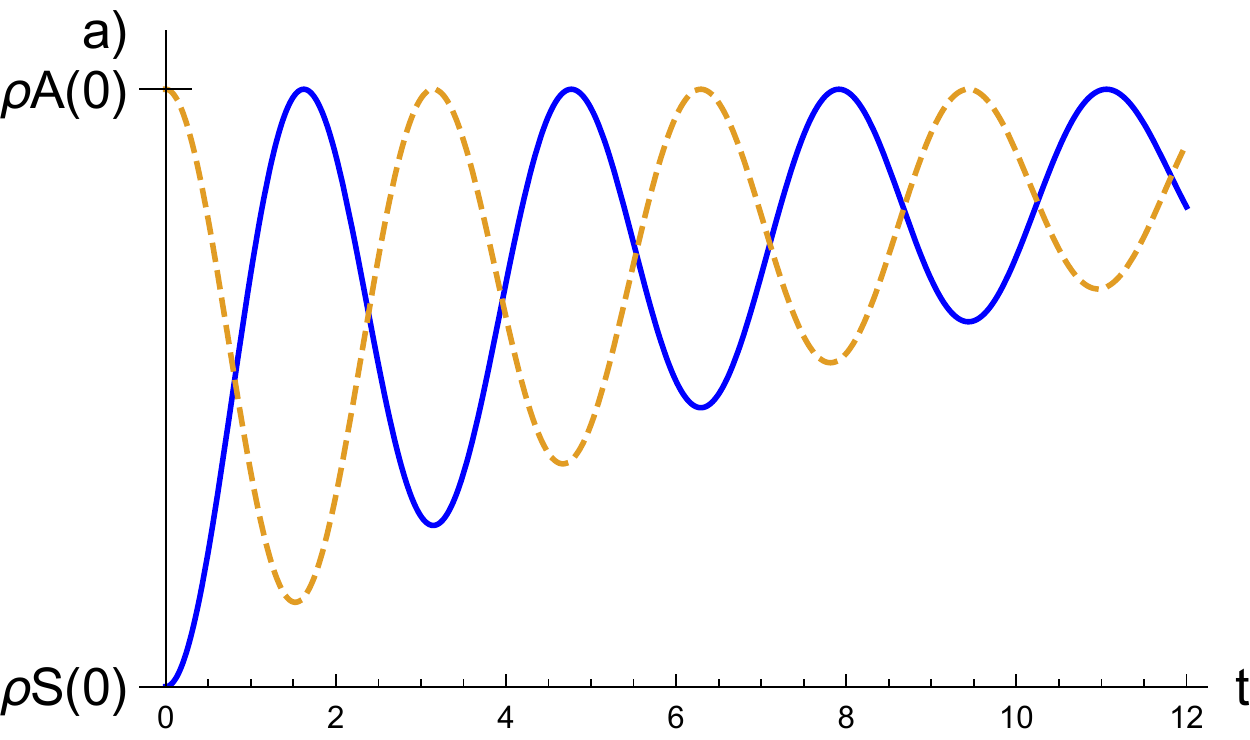}
\includegraphics[width=0.32\textwidth]{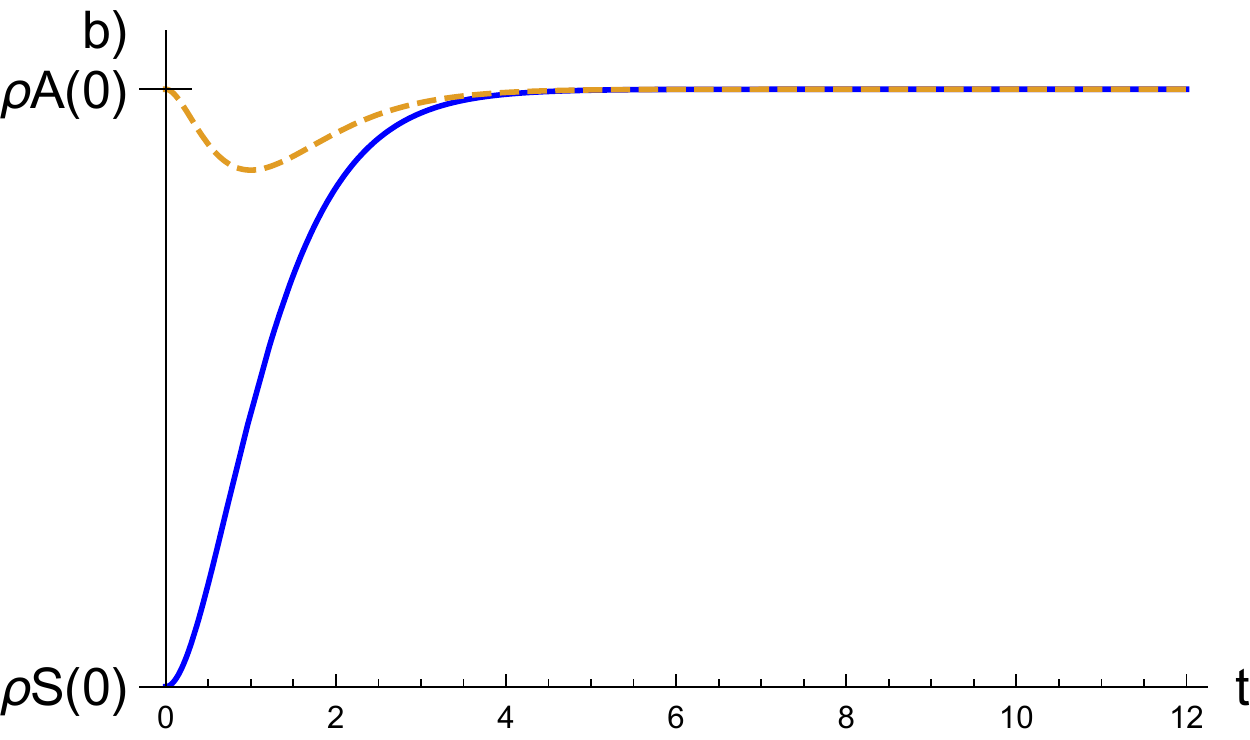}
\includegraphics[width=0.32\textwidth]{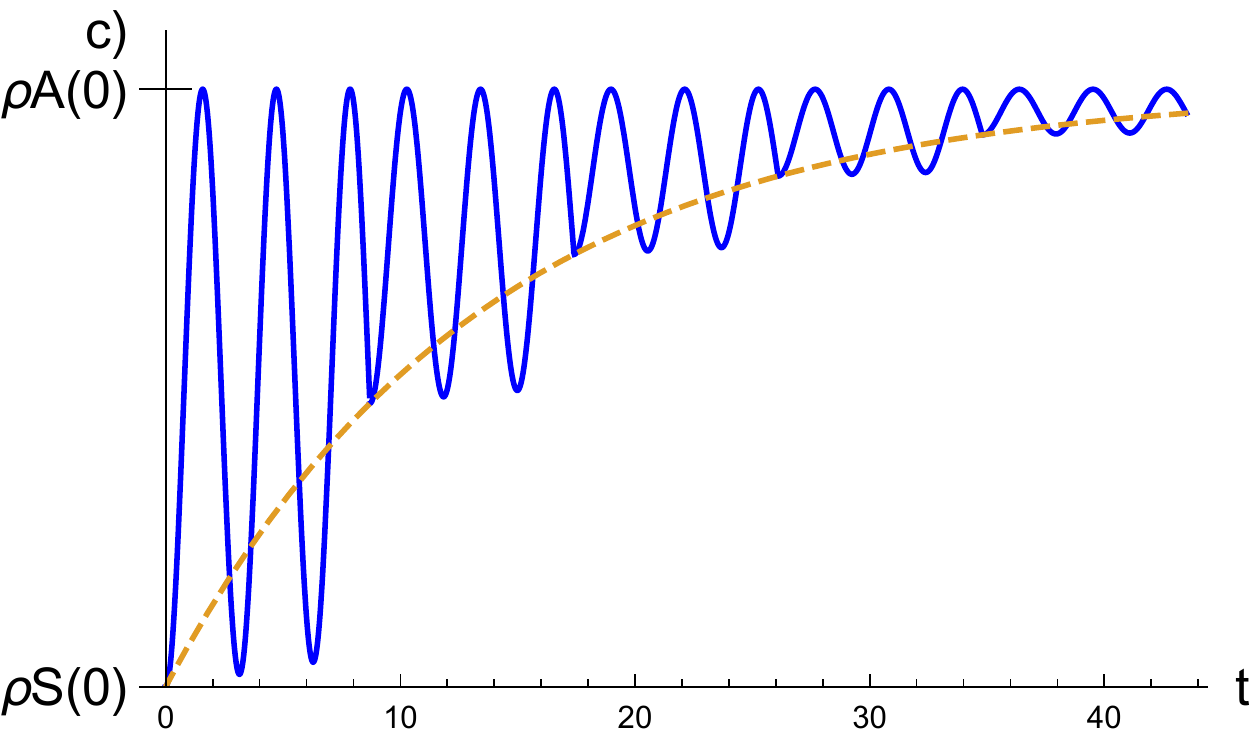}
\caption{(Color online.) The dynamics of two systems (S-solid, A-dashed) evolving under the damped partial swap interaction described in equation \eqref{DampSwapS} and \eqref{DampSwapA}. In figure a) the dynamics is under damped with $J=1$ and $\gamma_\text{A}=1/20$. In figure b) the evolution is critically damped with $J=1$ and $\gamma_\text{A}=1$. In figure c) the evolution of the system (solid) is tracked through its interactions with several ancillas ($J=1$, $\gamma_\text{A}=1/300$,  $\delta t=8.7$). Note that every $\delta t$ the system meets a new ancilla and begins oscillating between its new initial state and the initial state of the new ancilla. An exponential interpolation scheme (dashed) is also plotted. Note that the interpolation exactly matches the system state at the end of every interaction. }
\label{DampedFigure}
\end{figure*}

More realistically one might expect that during this interaction the ancilla is connected to a larger environment into which it leaks some information about the system's initial condition at some rate, $\gamma_\text{A}$. Using our harmonic oscillator analogy, one can imagine that the information about the system's initial state is stored in system A but dissipated into the environment  in the same way that the energy of a damped oscillator is stored as potential energy but dissipated while it is in motion
Motivated by this analogy, one can model the effect of A's environment by taking 
\begin{align}\label{DampSwapS}
\rho_\text{S}(t)
&= \phi_\text{S}(t) \ \rho_\text{S}(0)
+ (1-\phi_\text{S}(t)) \ \rho_\text{A}(0),\\
\label{DampSwapA}
\rho_\text{A}(t)
&=\phi_\text{A}(t) \, \rho_\text{A}(0)
+ (1-\phi_\text{A}(t)) \ \rho_\text{S}(0),
\end{align}
with 
\begin{align}\label{DampedPhiS}
\phi_\text{S}(t)
&=e^{-2\gamma_\text{A} t}\big( \cos(\omega \, t)
+\frac{\gamma_\text{A}}{\omega}
\sin(\omega\, t)
\big)^2,\\
\phi_\text{A}(t)
&=1-e^{-2\gamma_\text{A} t}\frac{J^2}{\omega^2}\sin(\omega \, t)^2,
\end{align}
and $\omega=\sqrt{J^2-\gamma_\text{A}^2}$ is the damped oscillation rate. Note that if $\gamma_\text{A}>J$ the oscillation is over-damped. Figure \ref{DampedFigure} a,b) shows the evolution of the two systems coupled this way when under damped and critically damped.

Alternatively, one could imagine that instead of swapping their initial states back and forth, the systems interact by repeatedly entangling and then disentangling. For example, the joint system could evolve as
\be
\rho_\text{SA}(t)
= \phi(t) \, \rho_\text{S}(0)\otimes\rho_\text{A}(0)
+ (1-\phi(t)) \, \ket{\psi}\bra{\psi},
\ee
where $\ket{\psi}$ is a maximally entangled state and $\phi(t)\in[0,1]$ describes the systems' evolution. In this case the dynamics of the systems' reduced states are,
\begin{align}
\rho_\text{S}(t)
&= \phi(t) \ \rho_\text{S}(0)
+ (1-\phi(t)) \ \hat{\openone}_\text{S}/D_\text{S},\\
\rho_\text{A}(t)
&=\phi(t) \, \rho_\text{A}(0)
+ (1-\phi(t)) \ \hat{\openone}_\text{A}/D_\text{A},
\end{align}
where $D_\text{S}$ and $D_\text{A}$ are the dimensions of the system and ancilla respectively. Again note that each system evolves within a one-dimensional space as a convex combination of two fixed endpoints.

We can capture the common elements of these examples in the following definition. In a \textit{one-dimensional convex collision model} the $n^{th}$ interaction updates the system and ancilla states as
\begin{align}\label{1DCCMUpdateS}
\rho_\text{S}(n \, \delta t)
&\to \phi_\text{S}(\delta t) \, \rho_\text{S}(n \, \delta t)
+(1-\phi_\text{S}(\delta t)) \, \rho_{\text{S},\odot},\\
\label{1DCCMUpdateA}
\rho_\text{A}(0)
&\to\phi_\text{A,n}(\delta t) \, \rho_\text{A}(0)
+(1-\phi_\text{A,n}(\delta t)) \, \rho_{\text{A},\odot,\text{n}},
\end{align}
for some $\phi_\text{S}(\delta t)$ and $\phi_\text{A,n}(\delta t)\in[0,1]$ and some target states $\rho_{\text{S},\odot}$ and $\rho_{\text{A},\odot,\text{n}}$. As in a generic collision model, we allow the details of the ancilla's update to depend on the interaction number, $n$, via a linear dependence on the current state of the system, $\rho_\text{S}(n\,\delta t)$. 

While in general both $\rho_{\text{A},\odot,\text{n}}$ and  $\phi_\text{A,n}(\delta t)$ can depend on $n$, we will now assume for simplicity that $\phi_\text{A,n}(\delta t)$ is independent of $n$. Note that this is the case in all of our motivational examples.

\subsection{Friction in one-dimensional convex collision models}
We will now calculate the average friction during the $n^{th}$ interaction, $f_n$, for a generic one-dimensional convex collision model.

First we note that the system's update equation, \eqref{1DCCMUpdateS}, can be easily solved yielding,
\be\label{PhiSSolution}
\rho_\text{S}(n\,\delta t)
=\phi_\text{S}(\delta t)^n \, \rho_\text{S}(0)
+(1-\phi_\text{S}(\delta t)^n) \, \rho_{\text{S},\odot}.
\ee
Next, we note that there is a natural interpolation scheme between the discrete time steps, $t=n\,\delta t$, given by,
\begin{align}\label{InterpolateFric}
\rho_\text{S}(t)
&=e^{-\Gamma \, t}\rho_\text{S}(0)
+(1-e^{-\Gamma \, t}) \ \rho_{\text{S},\odot},
\end{align}
where 
\bel{GammaDef}
\Gamma\coloneqq - \frac{1}{\delta t}\text{Ln}(\phi_\text{S}(\delta t)).
\ee
See Fig \ref{DampedFigure} c) for an illustration of such an interpolation scheme. Note that the interpolation scheme exactly matches the system state at the end of every interaction. Furthermore, if $\phi_\text{S}(\delta t)=0$ (such that the system reaches its target state after just one interaction and stays there) then $\Gamma=\infty$. Thus in this case the interpolation scheme predicts system reaches its target state just after $t=0$ and stays there.

Recall that the dependence on $n$ of the ancilla's target state is assumed to come from a linear dependence on $\rho_\text{S}(n\,\delta t)$. Using this linearity we know that the ancilla's target state must evolve as
\begin{align}\label{InterpolateFricA}
\rho_{\text{A},\odot,\text{n}}
=\phi_\text{S}(\delta t)^n \, \rho_{\text{A},\odot,0}
+(1-\phi_\text{S}(\delta t)^n)\rho_{\text{A},\odot,\infty}. 
\end{align}

Next, from equation \eqref{PhiSSolution} we can compute the system's internal energy at $t=n\,\delta t$ as,
\be
E_\text{S}(n\,\delta t)
=\phi_\text{S}(\delta t)^n \, E_\text{S}(0)
+(1-\phi_\text{S}(\delta t)^n) \, E_{\text{S},\odot},
\ee
where $E_\text{S}(0)=\text{Tr}_\text{S}(\hat{H}_\text{S}\, \rho_\text{S}(0))$ is the system's initial energy and $E_{\text{S},\odot}=\text{Tr}_\text{S}(\hat{H}_\text{S} \, \rho_{\text{S},\odot})$ is the energy of the system's target state. From this we can compute the change in the system's energy during the $n^{th}$ interaction,
\begin{align}
\delta E_\text{S,n}
&=(1-\phi_\text{S}(\delta t))\,
\phi_\text{S}(\delta t)^n\, (E_{\text{S},\odot}-E_\text{S}(0)).
\end{align}
Note that this is just a geometric sequence with a common ratio $\phi_\text{S}(\delta t)$ and normalized to have a sum of \mbox{ $E_{\text{S},\odot}-E_\text{S}(0)$}.

Similarly we can calculate that after the $n^{th}$ interaction the energy of the $n^{th}$ ancilla is,
\be
E_\text{A,n}
=\phi_\text{A}(\delta t) \, E_\text{A}(0)
+(1-\phi_\text{A}(\delta t)) \, E_{\text{A},\odot,\text{n}}.
\ee
where \mbox{$E_\text{A}(0)=\text{Tr}_\text{S}(\hat{H}_\text{S} \, \rho_\text{A}(0))$} is the energy of the ancilla's initial state and \mbox{$E_{\text{A},\odot,\text{n}}=\text{Tr}_\text{A}(\hat{H}_\text{A} \, \rho_{\text{A},\odot,\text{n}})$} is the energy of the $n^{th}$ ancilla's target state. The change in the ancilla's energy due to this interaction is,
\be
\delta E_\text{A,n}
=(1-\phi_\text{A}(\delta t)) \ (E_{\text{A},\odot,\text{n}}-E_\text{A}(0)).
\ee
From these we find that the friction averaged over the $n^{th}$ interaction is
\begin{align}
f_n&=(1-\phi_\text{S}(\delta t)) \ 
\phi_\text{S}(\delta t)^n \ 
\frac{E_{\text{S},\odot}-E_\text{S}(0)}{\delta x}\\
\nonumber
&+(1-\phi_\text{A}(\delta t)) \ \frac{E_{\text{A},\odot,\text{n}}-E_\text{A}(0)}{\delta x}.
\end{align}
That is, in the $n^{th}$ interaction the system takes an (ever diminishing) step towards its target state while the $n^{th}$ ancilla takes it first (and only) step towards its target state. 

We will now separate this friction into  permanent/transient parts that remain/vanish as $n\to\infty$. Specifically, we find 
\begin{align}\label{finfty}
f_\infty=(1-\phi_\text{A}(\delta t)) \ \frac{E_{\text{A},\odot,\infty}-E_\text{A}(0)}{\delta x}
\end{align}
for  the permanent friction.

Note that at late times the system has always reached its target state so the only energy cost is moving each ancilla one step towards its target state at $n=\infty$. Thus the permanent friction depends only on the dynamics of the ancillas.

The transient part of the friction is defined as
\begin{align}\label{ftrann}
f_\text{transient,n}
&\coloneqq f_n-f_\infty\\
\nonumber
&=(1-\phi_\text{S}(\delta t)) \ 
\phi_\text{S}(\delta t)^n \ 
\frac{E_{\text{S},\odot}-E_\text{S}(0)}{\delta x}\\
\nonumber
&+(1-\phi_\text{A}(\delta t)) \ \frac{E_{\text{A},\odot,\text{n}}-E_{\text{A},\odot,\infty}}{\delta x}.
\end{align}
The first term in the expression is associated with the system approaching its target state, $\rho_\text{S}(0)\to\rho_{\text{S},\odot}$. Similarly the second term is associated with the ancilla's target state approaching its final target state, $\rho_{\text{A},\odot,0}\to\rho_{\text{A},\odot,\infty}$.  From \eqref{InterpolateFricA} and the linear dependence of $E_{\text{A},\odot,\text{n}}$ on $\rho_{\text{A},\odot,\text{n}}$ we have
\begin{align}
E_{\text{A},\odot,\text{n}}-E_{\text{A},\odot,\infty}
=\phi_\text{S}(\delta t)^n \, (E_{\text{A},\odot,0}
-E_{\text{A},\odot,\infty}). 
\end{align}
Thus the second term in \eqref{ftrann} also decays geometrically at a rate $\phi_\text{S}(\delta t)$ each interaction. Factoring this decay out of both terms we find 
\be\label{ftrandecay}
f_\text{transient,n}
=f_\text{tr} \ \exp(-\Gamma \, n \, \delta t)
=f_\text{tr} \ \phi_\text{S}(\delta t)^n
\ee
where
\begin{align}
f_\text{tr}
\coloneqq f_\text{transient,0}
&=(1-\phi_\text{S}(\delta t)) \,
\frac{E_{\text{S},\odot}-E_\text{S}(0)}{\delta x}\\
\nonumber
&+(1-\phi_\text{A}(\delta t)) \, \frac{E_{\text{A},\odot,0}-E_{\text{A},\odot,\infty}}{\delta x}.
\end{align}
Thus   
\be
f_n=f_\infty+f_\text{tr} \ \exp(-\Gamma \, n \, \delta t)
\ee
and so $f_n$ is fully captured by the quantities, $f_\infty$, $f_\text{tr}$ and $\Gamma$.
Note that while we are characterizing the friction in terms of the interpolation parameter, $\Gamma$, our analysis only ever evaluated the states and energies of the systems at times, $t=n\,\delta t$, where the interpolation scheme is exact. The above equation can equivalently be interpreted as saying the transient friction decays geometrically by a factor of $\phi_\text{S}(\delta t)$ each interaction. The benefit of using the interpolation scheme is that it allows for fair comparisons of this decay for systems with different $\delta t$ (or equivalently travelling at different speeds). We will now make some general comments about each of these quantities.

First we note the the permanent friction, $f_\infty$, and the transient friction, $f_\text{tr}$ can both be either positive or negative depending on the energies of the system and ancilla's initial and target states. Specifically, we expect to see anti-friction when the energy of the system and ancilla's target state is lower than their initial state. As we will discuss later, such situations arise naturally from states with inverted populations. 

Next we note that $f_\infty$, $f_\text{tr}$, and $\Gamma$ can all depend on the systems' relative velocity through their dependence on $\delta t=\delta x/v$.

Finally, we note that the magnitude of the permanent and transient friction are both bounded as
\begin{align}
\vert f_\infty\vert
&\leq\frac{\vert E_{\text{A},\odot,\infty}-E_\text{A}(0)\vert}{\delta x},\\
\vert f_\text{tr}\vert
&\leq
\frac{\vert E_{\text{S},\odot}-E_\text{S}(0)\vert}{\delta x}
+\frac{\vert E_{\text{A},\odot,0}-E_{\text{A},\odot,\infty}\vert}{\delta x},
\end{align}
and therefore so is the total friction. Note that bounds are velocity independent, such that this model cannot predict $f\sim v$ for all $v$.  However we can predict this velocity profile in the low velocity regime as we shall see.

As we discussed in Sec \ref{SecZenoFriction}, the friction at high velocities depends on how the systems' update maps behave for small $\delta t$. For instance, suppose our regularity assumptions, \eqref{Regularity1} and \eqref{Regularity2}, are satisfied such that we can expand $\phi_\text{S}(\delta t)$ and $\phi_\text{A}(\delta t)$ around $\delta t=0$ as,
\begin{align}
\phi_\text{S}(\delta t)
&=1-\delta t \, \phi_\text{S,1}
+\mathcal{O}(\delta t^2),\\
\phi_\text{A}(\delta t)
&=1-\delta t \, \phi_\text{A,1}
+\mathcal{O}(\delta t^2),
\end{align}
then for large velocities we can expand the friction parameters as,
\begin{align}
f_\infty(v)
&=\frac{E_{\text{A},\odot,\infty}-E_\text{A}(0)}{v}\,\phi_\text{A,1}
+\mathcal{O}(v^{-2}),\\
\nonumber
f_\text{tr}(v)
&=\frac{E_{\text{S},\odot}-E_\text{S}(0)}{v}\phi_\text{S,1}
+\frac{E_{\text{A},\odot,0}-E_{\text{A},\odot,\infty}}{v}\phi_\text{A,1}+\mathcal{O}(v^{-2}),\\
\nonumber
\Gamma(v)&=\phi_\text{S,1}+\mathcal{O}(v^{-1}).
\end{align}
Note that as expected the magnitude of the friction goes as $1/v$ for large $v$, that is we see Zeno Friction. 

If we do not meet these regularity assumptions then at high velocities we will not see Zeno friction. For instance, if \mbox{$\phi_\text{S}(\delta t\to0)=1-F_S$} and \mbox{$\phi_\text{A}(\delta t\to0)=1-F_A$} then 
\begin{align}
f_\infty(v\to\infty)
&=\frac{E_{\text{A},\odot,\infty}-E_\text{A}(0)}{\delta x} F_A,\\
\nonumber
f_\text{tr}(v\to\infty)
&=\frac{E_{\text{S},\odot}-E_\text{S}(0)}{\delta x} \, F_S
+\frac{E_{\text{A},\odot,0}-E_{\text{A},\odot,\infty}}{\delta x} \, F_A\\
\nonumber
\Gamma(v)&=\text{Ln}(F_S) \, \frac{v}{\delta x}+\mathcal{O}(1),
\end{align}
for large $v$. That is, the permanent and transient friction both approach   constant values at high speeds, although the transient friction will vanish quickly since the decay rate becomes large.

The friction at low velocities depends on  how the system and ancillas interact for long times. 
For instance, if \mbox{$\phi_\text{S}(\delta t\to\infty)=1-f_S$} and \mbox{$\phi_\text{A}(\delta t\to\infty)=1-f_A$} then 
\begin{align}
f_\infty(v\to0)
&=\frac{E_{\text{A},\odot,\infty}-E_\text{A}(0)}{\delta x} f_A,\\
\nonumber
f_\text{tr}(v\to0)
&=\frac{E_{\text{S},\odot}-E_\text{S}(0)}{\delta x} \, f_S
+\frac{E_{\text{A},\odot,0}-E_{\text{A},\odot,\infty}}{\delta x} \, f_A\\
\nonumber
\Gamma(v\to0)&=0.
\end{align}
That is, the permanent and transient friction both approaches a constant at zero speeds. Note that since the decay rate goes to zero, so the transient friction will vanish very slowly.

If instead $\phi_\text{S}(\delta t)$ and $\phi_\text{A}(\delta t)$ decay polynomially to $1$ for large $\delta t$ as,
\begin{align}
\phi_\text{S}(\delta t)
&=1-\delta t^{-p} \, \phi_\text{S,p}\\
\phi_\text{A}(\delta t)
&=1-\delta t^{-p} \, \phi_\text{A,p},    
\end{align}
for some $p>0$ then we find for small velocities,
\begin{align}
f_\infty(v)
&=\frac{E_{\text{A},\odot,\infty}-E_\text{A}(0)}{\delta x^{p+1}} \, \phi_\text{A,p} \, v^p\\
\nonumber
f_\text{tr}(v)
&=\frac{E_{\text{S,t}}-E_\text{S}(0)}{\delta x^{p+1}} \, \phi_\text{S,p} \, v^p+\frac{E_\text{A,t,0}-E_{\text{A},\odot,\infty}}{\delta x^{p+1}} \, \phi_\text{A,p} \, v^p\\
\nonumber
\Gamma(v)&=
\frac{\phi_{S,-p}}{\delta x^{p+1}}v^{p+1}.
\end{align}
Thus we can recover any scaling behavior for small velocities by picking an appropriate exponent, $p$. 

\section{Examples}

We will now consider several example scenarios.
\subsection{Damped Partial Swap Interaction}\label{Sec:DampedPartialSwapExample}
Consider a spin qubit, S, moving at a speed $v$ relative to a line of spin qubit ancillas. Suppose that the ancillas are separated by a distance $\delta x$ and that the system interacts only with the nearest ancilla such that it meets a new ancilla, A, every $\delta t=\delta x/v$. Suppose that the system and ancillas are initially in thermal states,
\begin{align}
&\rho_\text{S}(0)=(1+a_\text{S}(0)\sigma_z)/2,
\qquad
\hat{H}_\text{S}=\hbar\omega_S\, \sigma_z/2,\\
&\rho_\text{A}(0)=(1+a_\text{A}(0)\sigma_z)/2,
\qquad
\hat{H}_\text{A}=\hbar\omega_\text{A}\, \sigma_z/2,
\end{align}
with respect to their local Hamiltonians. Note that for either system ($X=S,A$), $a_X=-1$ corresponds to the ground state ($n=0$) with the system's temperature increasing as $a$ increases. At $a_X=0$ the system is at infinite temperature, maximally mixed ($n=1/2$). For $a_X>0$ the state has an inverted population ($n>1/2$). 

Suppose the system couples to each ancilla via the isotropic spin coupling, $\hat{H}_\text{SA}=\hbar \, J \  \bm{\hat{\sigma}}_\text{S}\cdot
\bm{\hat{\sigma}}_\text{A}$. As discussed in Section \ref{MotivatingExamples} this coupling induces a partial swap interaction between the systems and corresponds to the one-dimensional convex collision model, \eqref{1DCCMUpdateS}, with
\begin{align}
\phi_\text{S}(\delta t)&=\cos(J \, \delta t)^2,\qquad
\rho_{\text{S},\odot}=\rho_\text{A}(0),\\
\phi_\text{A}(\delta t)&=\cos(J \, \delta t)^2,\qquad
\rho_{\text{A},\odot,\text{n}}=\rho_\text{S}(n\,\delta t).
\end{align}
This situation can be modified to include each ancilla dissipating information into its environment at a rate $\gamma_\text{A}$ by instead taking $\phi_\text{S}(\delta t)$ and $\phi_\text{A}(\delta t)$ to be
\begin{align}
\phi_\text{S}(\delta t)
&=e^{-2\gamma_\text{A} \delta t}\big( \cos(\omega \, \delta t)
+\frac{\gamma_\text{A}}{\omega}
\sin(\omega\, \delta t)
\big)^2,\\
\phi_\text{A}(\delta t)
&=1-e^{-2\gamma_\text{A} \delta t}\frac{J^2}{\omega^2}\sin(\omega \, \delta t)^2,
\end{align}
where $\omega=\sqrt{J^2-\gamma_\text{A}^2}$ is the damped oscillation rate. If the oscillation is overdamped, $\gamma_\text{A}>J$, then $\omega$ is imaginary. The identities, $\text{cos}(\ii x)=\text{cosh}(x)$ and $\text{sin}(\ii x)=\ii\,\text{sinh}(x)$ are useful in this case.

Computing the average friction during the $n^{th}$ interaction we find, $f_n=f_\infty+f_\text{tr} \, \exp(-\Gamma \, n \, \delta t)$ where,
\begin{widetext}
\begin{align}\label{DampedSwapResult}
f_\infty(v)&=0\\
f_\text{tr}(v)
\nonumber
&=\Bigg(\hbar\omega_S\bigg(1-e^{-2\gamma_\text{A} \delta x/v}\,\Big(\!\cos\Big(\frac{\omega \, \delta x}{v}\Big)
+\frac{\gamma_\text{A}}{\omega}
\sin\Big(\frac{\omega \, \delta x}{v}\Big)
\Big)^2\bigg)\,
-\hbar\omega_A\, \frac{J^2}{\omega^2}e^{-2\gamma_\text{A} \delta x/v} \sin\Big(\frac{\omega \, \delta x}{v}\Big)^2
\Bigg)\frac{a_A(0)-a_S(0)}{\delta x}\\
\Gamma(v)
\nonumber
&=2\gamma_\text{A}-\frac{2\, v}{\delta x}\text{Ln}\Big(\,\Big\vert \cos\Big(\frac{\omega \, \delta x}{v}\Big)
+\frac{\gamma_\text{A}}{\omega}
\sin\Big(\frac{\omega \, \delta x}{v}
\Big)\Big\vert\,\Big)\,.
\end{align}
\end{widetext}
Note that the friction is entirely transient. This is because at late times the system has reached its target state, \mbox{$\rho_\text{S}(\infty)=\rho_{\text{S},\odot}=\rho_\text{A}(0)$}, which is the ancilla's initial state. Thus at late times the partial swap interaction does not affect the reduced state of either system.
\begin{figure*}
\includegraphics[width=0.45\textwidth]{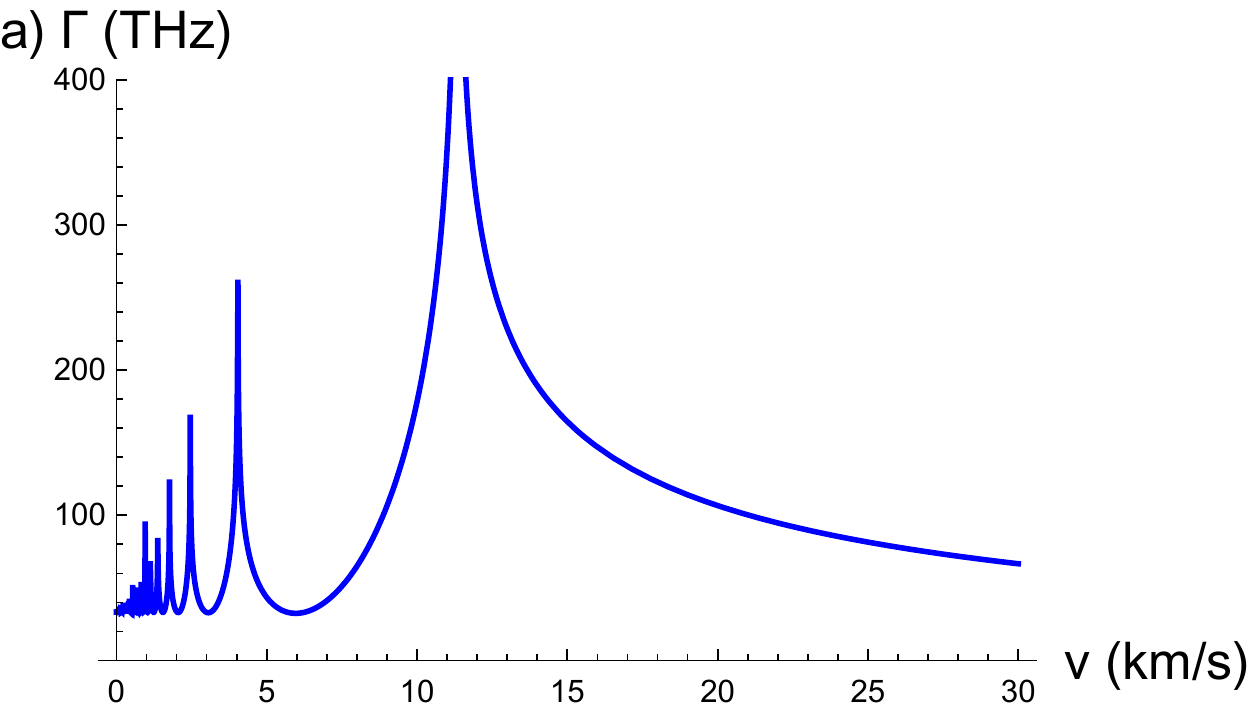}
\includegraphics[width=0.45\textwidth]{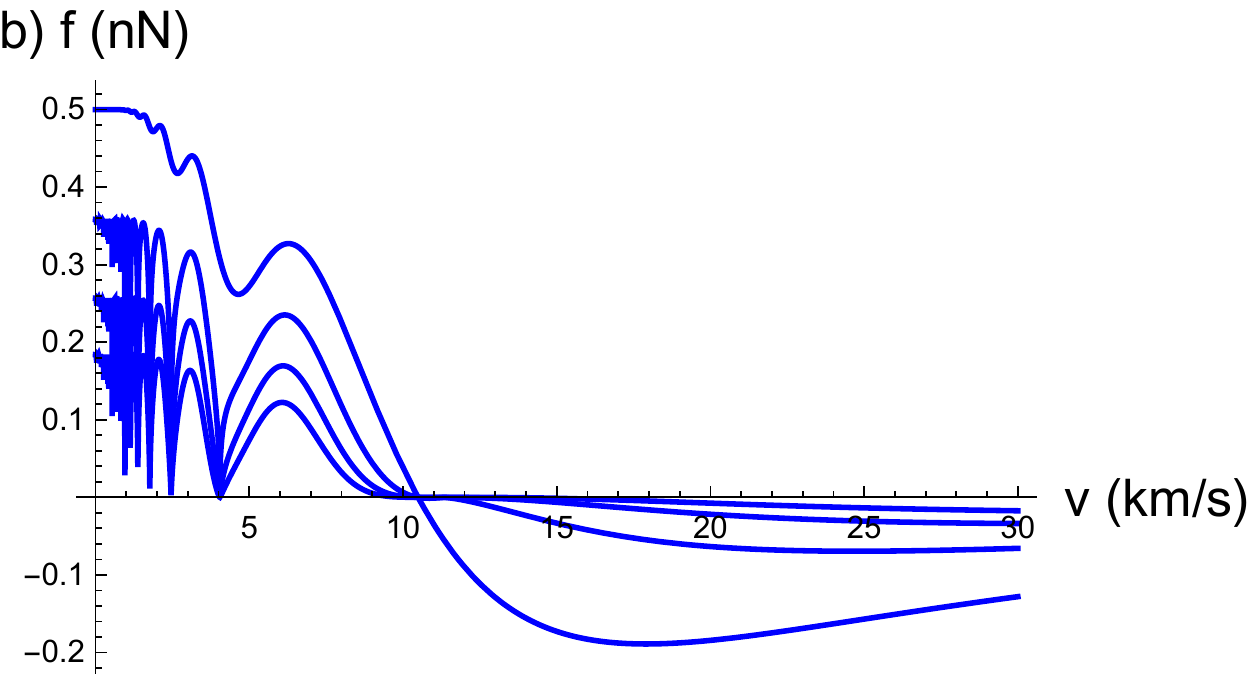}
\caption{(Color online.) The friction profile of a spin qubit, S, moving over a surface of ancillary spin qubits, A, spaced a distance $\delta x=0.2$  nm from each other. The system is initially in its maximally mixed state, $a_\text{S}(0)=0$, with an internal energy scale, $\hbar\omega_\text{S}=0.6$ eV. 
The ancillas are each initially in their excited state, $a_\text{A}(0) = 1$, and have an internal energy scale, $\hbar\omega_\text{A}=1.2$ eV. The systems couple via an isotropic spin coupling with $\hbar J=60$ meV and an ancilla leaks information into its environment at a rate $\gamma_\text{A}=16$ THz.  The evolution is underdamped with a frequency $\omega=100$ THz. At the left, in diagram (a) we depict the velocity dependent decay rate of the friction. Note the divergences when $\omega\delta x/v\sim (n+1/2)\pi$, and the ``baseline'' decay rate of $2\gamma_\text{A}=33$ THz. In (b) the velocity dependence of friction is shown at times $t=0,\,10,\,20,\,30$ fs as it decreases towards zero.}
\label{DampedExamples}
\end{figure*}

For large velocities we can expand the friction as a series in $1/v$ to find,
\begin{align}\label{DampedPartialSwapFtrGamma}
\nonumber
f_\text{tr}(v)
&=\frac{\hbar \, J^2 \, \delta x}{v^2}
(\omega_\text{S}-\omega_\text{A})(a_\text{A}(0)-a_\text{S}(0))
+\mathcal{O}(v^{-3})\\
\Gamma(v)
&=\frac{J^2 \, \delta x}{v} +\mathcal{O}(v^{-2}).
\end{align}
We note that these expansions hold independently of whether the evolution is over-, under-, or critically damped. As predicted in Sec \ref{SecZenoFriction}, the magnitude of the friction goes to zero as the velocity increases. 

We see from \eqref{DampedPartialSwapFtrGamma} that the transient friction at large velocities can be either positive or negative depending on the relative energy gaps and  initial polarizations.  Most strikingly, we see anti-friction at high velocities when the system with the higher energy gap also has a higher population number.

Taking the limit of small velocities we find
\begin{align}
f_\text{tr}(v\to0)
&=\frac{\hbar\omega_\text{S}}{\delta x} (a_\text{A}(0)-a_\text{S}(0)),\\
\Gamma(v\to0)
&=\begin{cases}
2\gamma_A, &\gamma_A\leq J\\
2\gamma_\text{A}-2\sqrt{\gamma_\text{A}^2-J^2}, &\gamma_A> J
\end{cases}
\end{align}
and once again the transient friction as $v\to0$ can be either positive or negative depending on the  initial polarizations. At low velocities anti-friction is clearly manifest when the system is more populated than the ancillas. This is the can be the case (for instance) if the ancillas are all in their ground state.

At intermediate velocities the transient friction can oscillate and change sign as shown in Fig \ref{DampedExamples}. 

In this and all following figures we have picked our dimensionful quantities along the lines of the Van der Waals interaction example discussed above after \eqref{ZenoCriticalSpeed}. For reference, a force of $F=0.1$ nN acting on a nitrogen atom with mass $m=14$ amu results in an acceleration of $4.3\times10^{15} \ \text{m}/\text{s}^2$. From an initial speed of $10$ km/s, this force stops the atom in $2.3$ ps over a distance of $11$ nm. Travelling this distance, the atom would cross $50$ Van der Walls radii.

We can modify this scenario to avoid Zeno friction (i.e., to have friction at large velocities) by having a velocity dependent coupling. For example we could take the coupling strength $J$ to be velocity dependent, with $J=k\, v$ for some $k$. Calculating the friction in this case yield the same result  \eqref{DampedSwapResult} as before, but with $\omega=\sqrt{k^2v^2-\gamma_\text{A}^2}$. Note that for large velocities the dynamics is always underdamped. Likewise for small velocities the dynamics is always over damped.

For large velocities, we can expand the transient friction and decay rate 
\begin{align}
f_\text{tr}(v)
&=\frac{\hbar\,\sin(k \, \delta x)^2}{\delta x}(\omega_\text{S}-\omega_\text{A})(a_\text{A}(0)-a_\text{S}(0))+\mathcal{O}(v^{-1})\\
\Gamma(v)
&=\frac{v}{\delta x} \text{Log}(\sec(k\,\delta x)^2)
+\mathcal{O}(1).
\end{align}
and, as anticipated, in this case the friction does not decay as $1/v$ for large $v$. However since the  decay rate 
$\Gamma$ is proportional to $v$, the friction at high velocities decays quickly. Anti-friction will take place at high velocities when the system with the higher energy gap also has a higher population number.

For small velocities we obtain
\begin{align}
f_\text{tr}(v)
&=\frac{k^2 \, v \, \hbar\omega_\text{S}}{\gamma_\text{A}}
(a_\text{A}(0)-a_\text{S}(0))+\mathcal{O}(v^2)\\
\Gamma(v)
&=\frac{k^2 v^2}{\gamma_\text{A}}+\mathcal{O}(v^3)
\end{align}
upon expansion.  In this regime we recover the usual friction dependence $f\sim v$. Moreover note that in this regime the friction's decay rate is very small, meaning that while the friction is entirely transient it will last a relatively long time. As in the previous example, we see anti-friction when the system has a higher population number than the ancillas. 

\begin{figure*}
\includegraphics[width=0.45\textwidth]{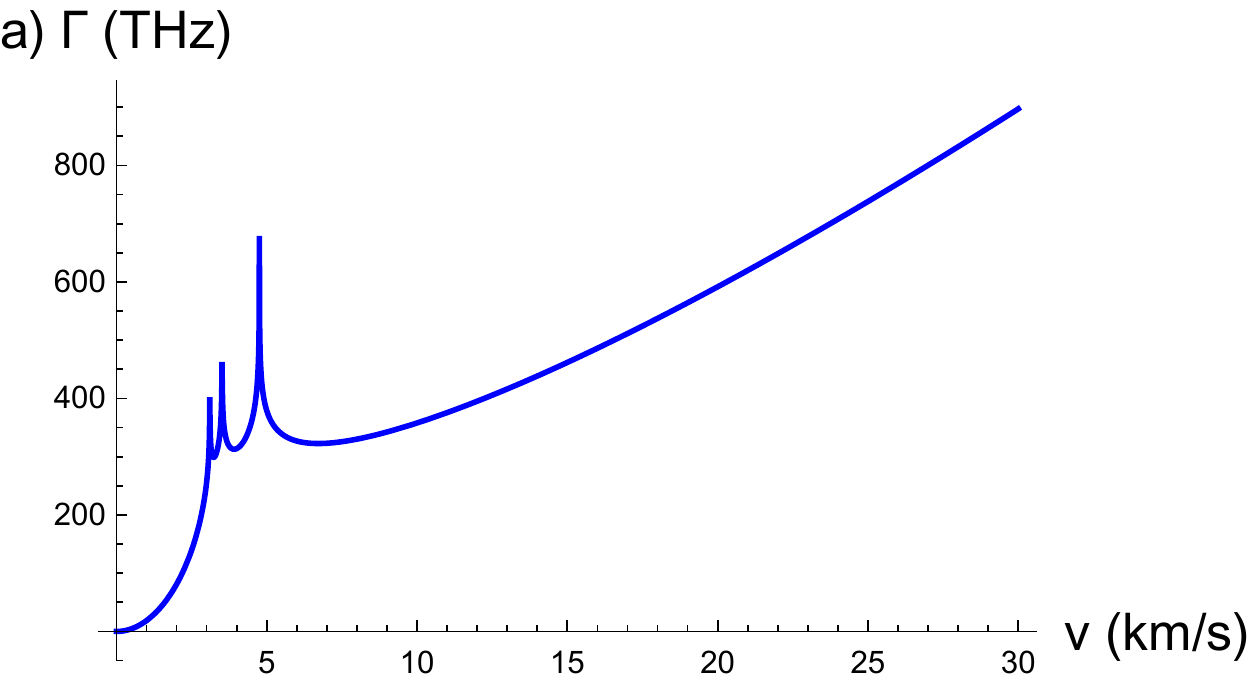}
\includegraphics[width=0.45\textwidth]{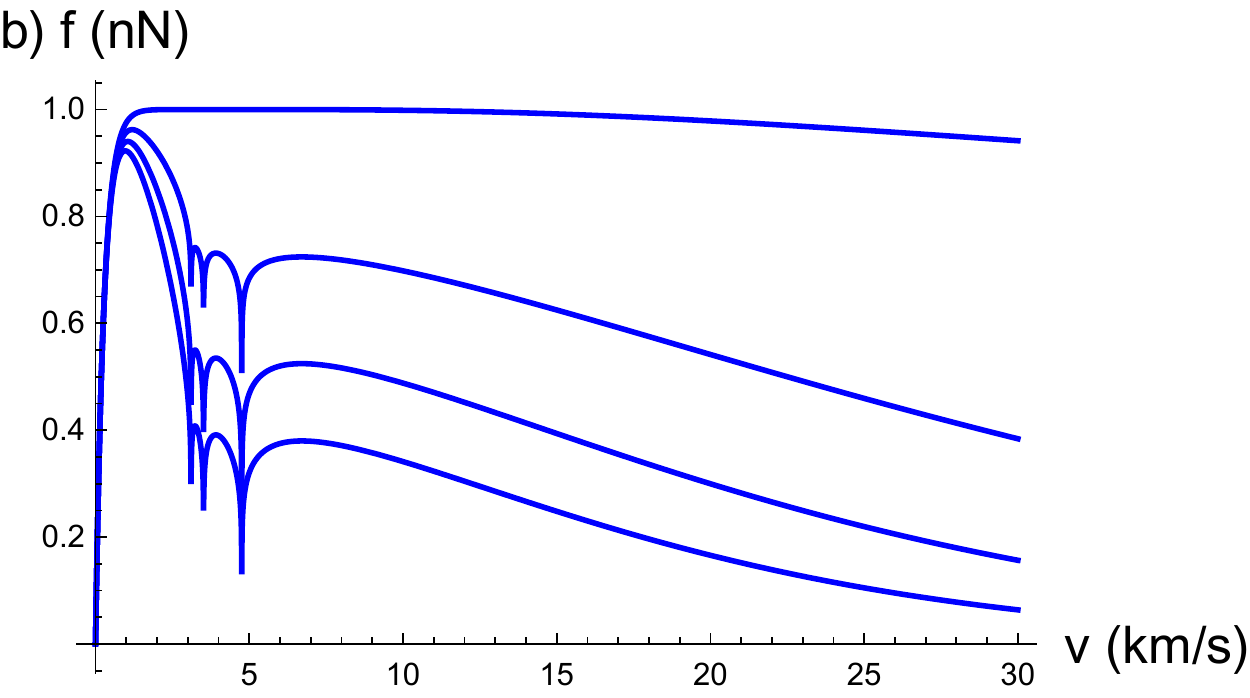}
\caption{(Color online.) The friction profile of a spin qubit, S, moving over a surface of ancillary spin qubits, A, spaced a distance $\delta x=0.2$  nm from each other. The system is initially in its ground state, $a_\text{S}(0)=-1$, with an internal energy scale, $\hbar\omega_\text{S}=1.2$ eV. 
The ancillas are each initially in their maximally mixed state, $a_\text{A}(0) = 0$, and have an internal energy scale, $\hbar\omega_\text{A}=0.6$ eV. The systems couple via an isotropic spin coupling with a velocity dependent coupling strength $J=k \, v$ with $k=55 \text{nm}^{-1}$. The ancillas leak information into their environment at a rate, $\gamma_\text{A}=165$ THz. The evolution is underdamped for large velocities, over damped for small velocities, and critically damped at  speed $v_c=\gamma_\text{A}/k=3$ km/s,  with a frequency $\omega=100$ THz. The velocity dependent decay rate of the friction is depicted in diagram (a) at the left, and the velocity dependence of the friction is shown at times $t=0,\,1,\,2,\,3$ fs at the right in (b) as it decreases towards zero.}
\label{DampedExamplesVDep}
\end{figure*}

Figure \ref{DampedExamplesVDep} shows the behavior of the friction at intermediate velocities. Note that the dynamics can be critically damped at intermediate velocities.

\subsection{Entangling-disentangling Interactions}\label{Sec:EDExample}

Next let us consider the second motivating example described in Sec. \ref{MotivatingExamples} in which the system and ancilla repeatedly entangle and disentangle with each other. As discussed above this dynamics can be described by the one-dimensional convex collision model with,
\begin{align}
\phi_\text{S}(\delta t)&=\epsilon+(1-\epsilon)\cos(J \, \delta t)^2,\qquad
\rho_{\text{S},\odot}=\hat{\openone}_\text{S}/D_\text{S},\\
\phi_\text{A}(\delta t)&=\epsilon+(1-\epsilon)\cos(J \, \delta t)^2,\qquad
\rho_{\text{A},\odot,\text{n}}=\hat{\openone}_\text{A}/D_\text{A}
\end{align}
for some $0\leq\epsilon\leq1$ and some oscillation rate $J$. Note that if $\epsilon\neq0$ then the system and ancilla are never maximally entangled with each other.

From these we compute
\begin{align}
f_\infty(v)&=(1-\epsilon)\sin\Big(\frac{J \, \delta x}{v}\Big)^2 \ 
\frac{E_{\text{A},\odot,\infty}-E_\text{A}(0)}{\delta x}\nonumber\\
f_\text{tr}(v)
\label{eq71}
&=(1-\epsilon)\sin\Big(\frac{J \, \delta x}{v}\Big)^2 \ 
\frac{E_{\text{S},\odot}-E_\text{S}(0)}{\delta x}\\
\Gamma(v)
\nonumber
&=\frac{v}{\delta x}\text{Ln}\Big(1-(1-\epsilon)\sin\Big(\frac{J \, \delta x}{v}\Big)^2\Big)
\end{align}
and find that at all velocities, the transient and permanent friction are each negative if and only if the systems' initial states are at higher energies than the maximally mixed states, or in other words when the population numbers of the state are skewed towards higher energies. In this case the presence of anti-friction is directly tied to inverted populations.

Expanding the quantities in \eqref{eq71} at large velocities yields
\begin{align}
f_\infty(v)&=(1-\epsilon)
\frac{J^2 \, \delta x^2}{v^2}
\frac{E_{\text{A},\odot,\infty}-E_\text{A}(0)}{\delta x}+\mathcal{O}(v^{-4})\nonumber \\
f_\text{tr}(v)
&=(1-\epsilon)
\frac{J^2 \, \delta x^2}{v^2}
\frac{E_{\text{S},\odot}-E_\text{S}(0)}{\delta x}+\mathcal{O}(v^{-4})\\
\Gamma(v)
\label{eq72}
&=\frac{J^2 \, \delta x}{v}(1-\epsilon)
+\mathcal{O}(v^{-3}).
\end{align}
and see that, as predicted by Section \ref{SecZenoFriction}, the friction decays towards zero for large enough velocities. 

\begin{figure*}
\includegraphics[width=0.45\textwidth]{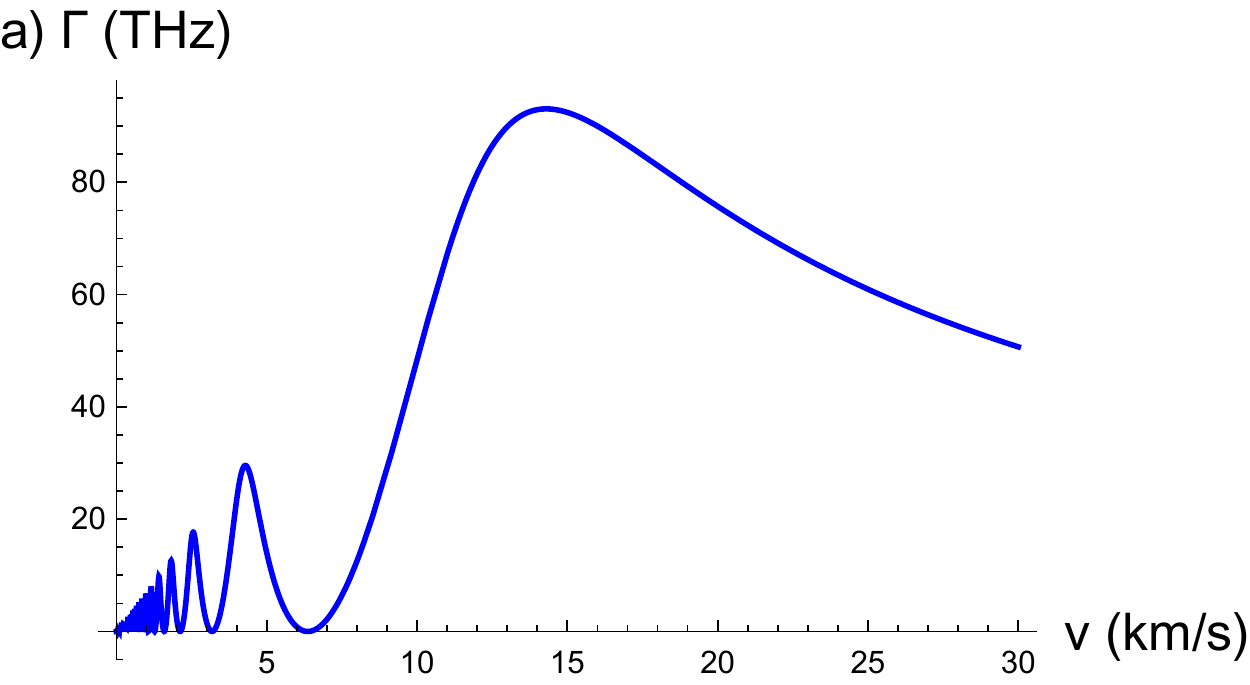}
\includegraphics[width=0.45\textwidth]{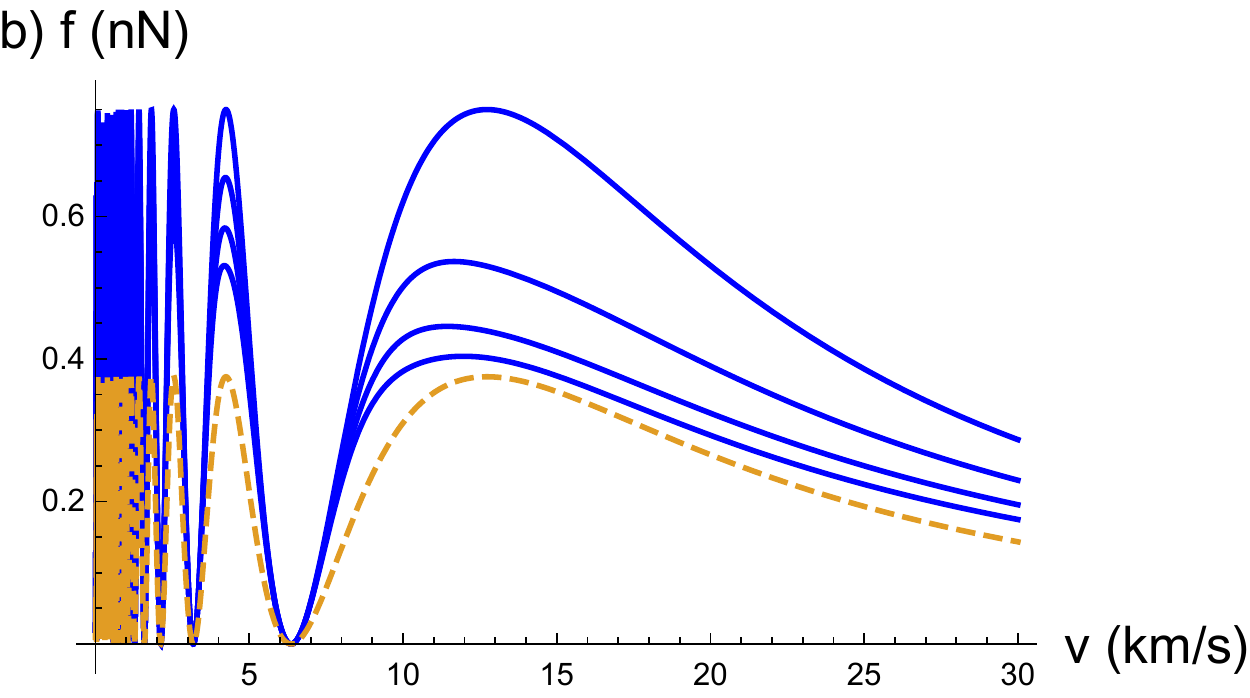}
\caption{(Color online.) The friction profile a system, S, moving over a surface of ancillary systems, A, spaced a distance $\delta x=0.2$ nm from each other. The system is initially in a state with energy $E_\text{S}(0)=0$ and evolves towards its maximally mixed state with energy $E_{\text{S},\odot}=0.6$ eV. The ancillas are each initially in a state with energy $E_\text{A}(0)=0$ and evolve towards their maximally mixed state with energy $E_{\text{A},\odot,\infty}=0.6$ eV. While interacting the systems entangle and disentangle with each other  at a rate $J=100$ THz and with $\epsilon=1/4$, as described in Section \ref{MotivatingExamples}.  At left, in diagram (a) we illustrate the velocity dependence of the decay rate of the friction, and in diagram (b) at right we show  the velocity dependence of the friction   at times $t=0,\,10,\,20,\,30$ fs from top to bottom (solid), as well as the friction at $t=\infty$ (dashed).}
\label{EDExample}
\end{figure*}

For small velocities, both the permanent and transient oscillate as $\text{sin}(1/v)^2$ and therefore do not converge as $v\to0$, though the decay rate does converge to zero: $\Gamma(v\to0)=0$.
Figure \ref{EDExample} shows the behavior of the friction at intermediate velocities.

\begin{figure*}
\includegraphics[width=0.45\textwidth]{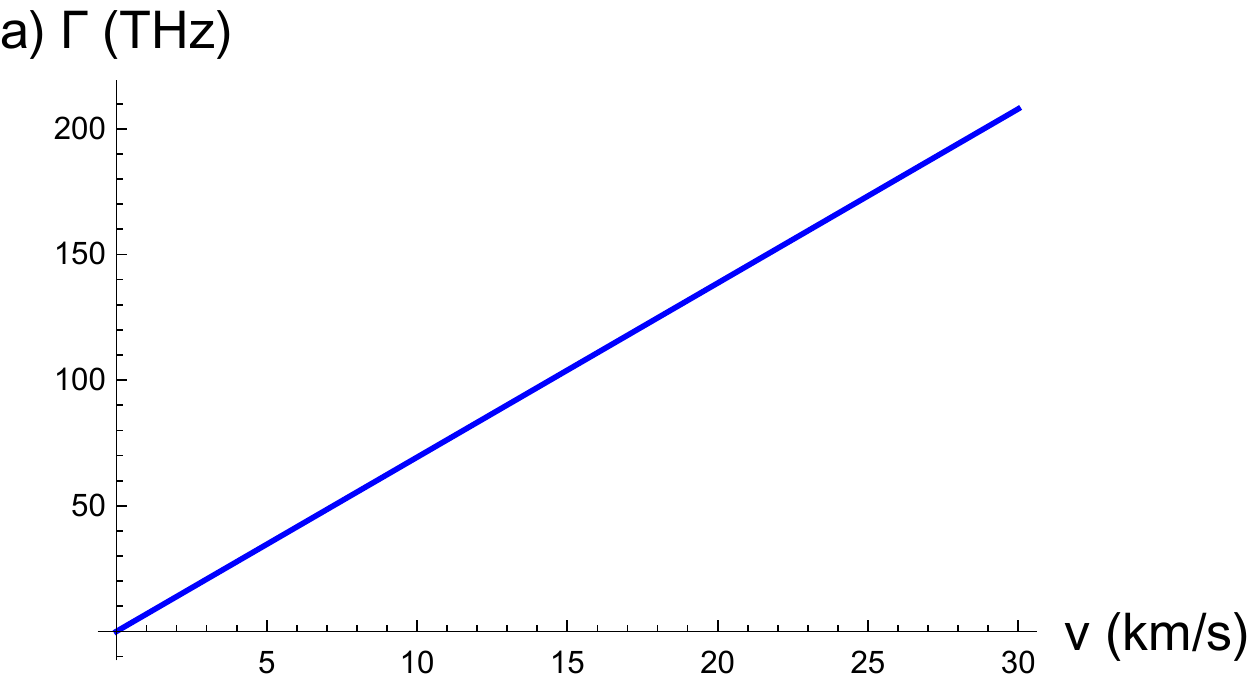}
\includegraphics[width=0.45\textwidth]{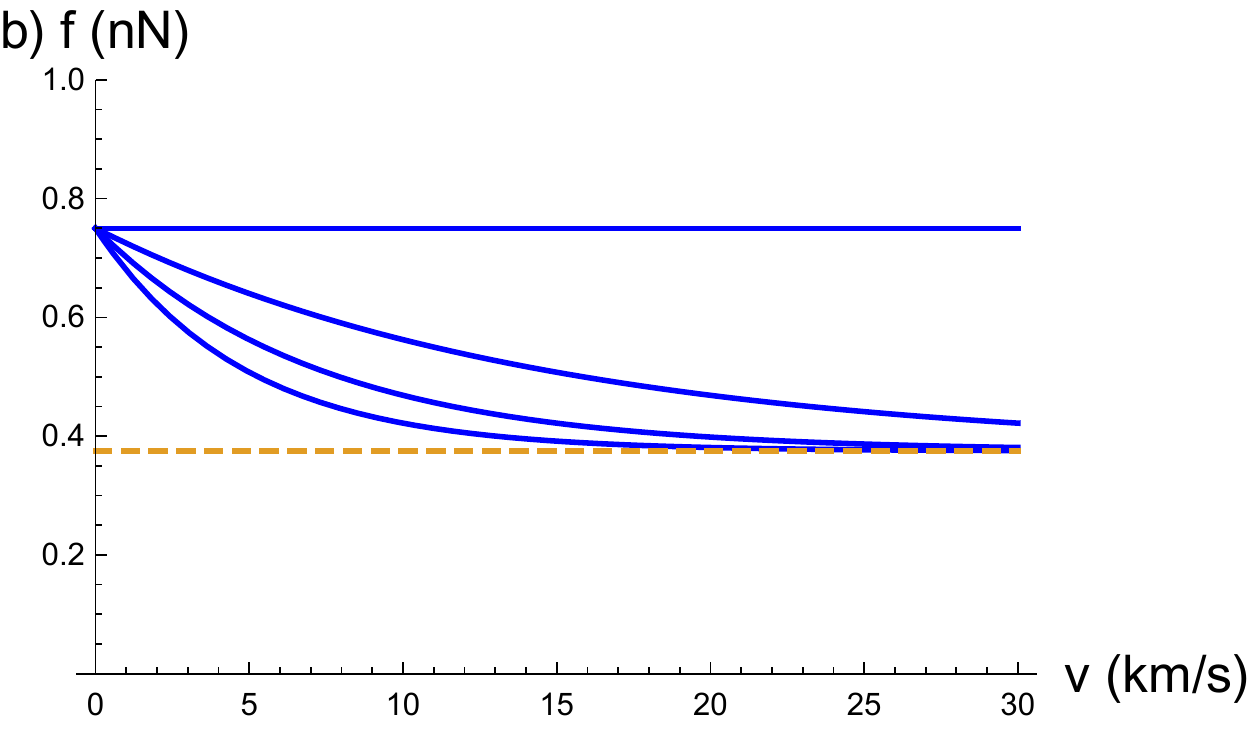}
\caption{(Color online.) We depict the friction profile of  system S moving over a surface of ancillary systems, A, spaced a distance $\delta x=0.2$ nm from each other. The system is initially in a state with energy $E_\text{S}(0)=0$ and evolves towards its maximally mixed state with energy $E_{\text{S},\odot}=0.6$ eV. The ancillas are each initially in a state with energy $E_\text{A}(0)=0$ and evolve towards their maximally mixed state with energy $E_{\text{A},\odot,\infty}=0.6$ eV. While interacting, the system and ancillas entangle and disentangle with each other  at a velocity dependent rate $J=k v$ with $k=55 \text{nm}^{-1}$ and with $\epsilon=1/4$. We exhibit in diagram (a) at left the velocity dependence of the decay rate of the friction and in diagram (b) at right the velocity dependence of the friction at times $t=0,\,10,\,20,\,30$ fs from top to bottom (solid) as well as the friction at $t=\infty$ (dashed).}
\label{EDExampleVDep}
\end{figure*}

We can modify this example by taking the coupling strength to be velocity dependent. For example, taking $J=k\,v$ for some $k$ we find
\begin{align}
f_\infty(v)&=
\sin\big(k\,\delta x\big)^2 \, (1-\epsilon)\,  \frac{E_\text{A,f}-E_\text{A}(0)}{\delta x}\\
f_\text{tr}(v)
&=\sin\big(k\,\delta x\big)^2 \, (1-\epsilon) \,  \frac{E_{\text{S},\odot}-E_\text{S}(0)}{\delta x}\\
\Gamma(v)
&=\frac{v}{\delta x}\text{Ln}\Big(1-(1-\epsilon)\sin\big(k \, \delta x\big)^2\Big)
\end{align}
where we see that $f_\infty$ and $f_\text{tr}$ do not depend on velocity. The decay rate  $\Gamma$ is now simply proportional the the systems' relative velocity. We plot this in Figure \ref{EDExampleVDep}. Note that at small velocities the transient friction decays very slowly.

\section{Discussion}

\subsection{Anti-friction in active/inverted media}
As we have shown, anti-friction arises in our model when the initial states of the system and ancillas have higher energies than their target states. As they approach their final states their internal energies are lowered. In our model the energy released by this process goes to the external agent controlling the motion of the systems. As we have argued, if the systems are moving under their own inertia, the excess energy will go into their kinetic energies, speeding them up. We note this conclusion may not hold if the interaction between the atom and the surface is mediated through some other system, say a quantum field. In this case, the excess energy could be absorbed by the field or carried off as radiation.

Under what conditions can we expect the systems' target states to be of higher energy than their initial states? In the example discussed in Sec. \ref{Sec:EDExample}, the systems' target states are both the maximally mixed states. Thus we saw anti-friction when the systems had inverted populations, that is with population distributions skewed towards higher energies. In this example, in order to have non-transient anti-friction the ancilla states must be inverted. This suggests that a particle travelling through an active media, such as lasing media, could be accelerated as it travels through it. Of course, as with lasing, this process would deplete the medium, which would have to be continuously repumped.

However, as we saw in section \ref{Sec:DampedPartialSwapExample} inverted populations are not necessary for anti-friction. In this example we saw that anti-friction at low velocities when the traveling spin-qubit had higher population than the ones composing the surface. This does not require inverted populations for either system. In fact, assuming that the system and ancillas do not have inverted populations we can write this condition in terms of their temperatures as $T_S/\omega_S>T_A/\omega_A$. That is, we see anti-friction as long as the system has a temperature higher than the ancillas times the ratio of their energy gaps. Note that this is always true when the ancillas are in their ground state! We predict that a finite temperature spin-qubit moving over a surface at $T=0$ will be accelerated by its interaction with the surface.  

This raises the question as to where is the energy coming from if the ancillas are all already in their lowest energy state. Noting that in this case the friction is entirely transient, we can see that the energy comes from the initial internal energy of the system, which it slowly converts to kinetic energy as it interacts with the ancillas. One can show that if the ancillas are all in their ground state, then the permanent friction must be positive, such that any anti-friction must be a transient phenomena. To see this note from equation \eqref{finfty} that if $E_A(0)$ is the lowest energy possible for the ancillas, then the permanent friction must be positive. 

Is it possible to have permanent anti-friction without having the ancillas in an inverted population? We leave this as an open question.

\subsection{Comparison with Casimir-type Quantum Friction}
In the introduction, we contrasted our approach to building a quantum model of friction with the Casimir-type quantum friction considered elsewhere. Namely, our method simplifies the scenario somewhat by not including the quantum field which mediates the interaction. However, our approach adds additional features to the neutral object, namely an internal structure. In the Casimir approach the neutral objects are treated as boundary conditions against which the quantum field theory equations are solved. That is they ignore\footnote{We do note that in \cite{PhysRevA.98.032507} and \cite{Barton_2010} the friction induced on an atom moving over a surface is computed, treating the atom as an Unruh-DeWitt detector or harmonic oscillator respectively, i.e. with internal structure. However, we note that in these works the surface is still treated as a boundary condition.} the internal structure of the macroscopic objects that set the boundary conditions.

While these two approaches are very different, they can produce similar scales of force and velocities. For instance in \cite{PhysRevA.98.032507} forces of $\sim 10^{-10}\text{ N}$ are seen at velocities of $10^{-4}c=30\text{ km/s}$ in line with the results of this paper.

Another point of comparison with these models is how the friction scales with velocity at low speeds. For instance, in \cite{Intravaia_2015} it was found that $f\sim v^3$. While we did not find this particular scaling in any of our examples,  our model can predict such scaling, as discussed above.

Additionally, it is worth discussing to what degree these approaches can model destructive friction. That is, friction which breaks or rearranges the bonds between the components of a rough surface. It is unclear how the Casimir-type could model destructive friction since it does not even consider the surface as being composed of bound parts.

It is easier to see how the collisional approach could model destructive friction; the energy cost of breaking these bonds could be included in the ancilla's energy. However, if the ancillas are sufficiently coupled to each other they may become correlated, violating our assumption that the ancilla states are independent.

\section{Conclusion}
We analyzed the friction induced on a quantum system as it moves over a surface composed of other quantum systems, i.e., we modeled both the system and the surface as quantum systems.  The interactions are described by a generic {collision model}, with the moving system interacting with the constituents of the surface one at a time.

With only mild regularity assumptions about their interaction, we found, unexpectedly, that the magnitude of the friction decays as $1/v$ for large enough velocities. We term this phenomena Zeno friction because it involves short interactions at a rapid rate.

To explore friction at low velocities and with velocity dependent couplings we motivated and developed what we call \textit{one-dimensional convex collision models}. These models include the ubiquitous partial swap interaction (which we show can be modified to include the surface dissipating into the bulk) as well as the system repeatedly entangling and disentangling with the constituents of the surface. We computed  the friction induced by such interactions in general, as well as these examples in particular.  

In general, the friction decays exponentially over time from its initial value $f(0,v)$ to its final value $f(\infty,v)$ at a some rate, $\Gamma(v)$. All three of these parameters can have a very complicated velocity dependence. Moreover the friction can be negative, indicating that the system accelerates due to its interaction with the surface. We found that this phenomena is associated with a population inversion of the constituents of the surface. In addition to these extreme possibilities, we also recovered standard friction profiles ($f(v)\sim v$ and $f(v)=\text{const}$) in certain low velocity scenarios.

\section*{Acknowledgements}
AL, RBM and EMM acknowledge support through the Discovery Grant Program of the Natural Sciences and Engineering Research Council of Canada (NSERC). EMM also acknowledges support through an Ontario ERA award. DG acknowledges support by NSERC through a Vanier Scholarship.

\appendix
\section{Is the energy cost of friction always converted to heat?}\label{Conservative}
It is interest to consider whether or not it is a necessary property of friction that its energy cost be converted into heat. For instance, one may ask, ``Does regenerative braking in cars count as friction?'' As a simpler example imagine a metal sphere dragged some distance across the carpet; Generally, the sphere will end up both hot and charged. That is, the energy cost is paid into both  heat (unrecoverable) and charge (recoverable), $W_\text{tot}=W_\text{heat}+W_\text{rec}$. Dividing this equation by the distance traveled, $\Delta x$ we can split the total motion-resisting force  into two parts, $f_\text{tot}=f_\text{heat}+f_\text{rec}$.

One may argue that, since the energy stored in the sphere's charge can be recovered, say by a controlled discharge, it should not count as friction. However  whether or not some store of energy is recoverable will depend on not only which energy recovery techniques are available but the frequency at which they are applied. For example, suppose that the charge on the sphere is discharged into its environment at some rate, $\Gamma_\text{dis}$, and that the energy released from these discharging events is converted into heat. If one performs controlled discharges on the sphere much less frequently than this rate, almost none of this charging energy is recovered. 
However if one discharges the sphere much more frequently than this rate, one can recover more of this energy cost before it becomes heat, yielding a lower friction. One could imagine that with very sophisticated intervention all of the friction could be eliminated. This context dependence could be anticipated by recalling that friction is fundamentally a phenomenon associated with open systems; if we can control all parts of our systems then all friction can be removed.

Alternatively one may argue that friction should be defined as the the total motion-resisting force that one has to pull against. With respect to the above discussion, this is equivalent to making no effort to recover any of the energy cost, or of assuming the energy decays to heat very quickly. In this paper we follow this latter suggestion. Note that this allows us to apply our analysis to generic scenarios where the ability-to-recover-energy of our agents is not specified.

\bibliography{references}

\end{document}